\documentclass[longauth]{aa} 

\usepackage{natbib}
\bibpunct{(}{)}{;}{a}{}{,} 

\usepackage{soul}
\usepackage{graphicx}
\usepackage{txfonts}
\usepackage{url}
\usepackage{color}
\usepackage{natbib}
\usepackage{rotating}
\usepackage{lineno} 
\usepackage{xspace}

\newcommand{\pmn}{PMN\,J1603$-$4904\xspace}

\bibpunct{(}{)}{;}{a}{}{,} 
\begin{document} 
\title{The unusual multiwavelength properties of the gamma-ray source
  PMN\,J1603$-$4904} 

\author{
  Cornelia M\"uller\inst{\ref{affil:remeis},\ref{affil:wuerzburg}}
  \and M.~Kadler\inst{\ref{affil:wuerzburg}}
  \and R.~Ojha\inst{\ref{affil:nasa_gfsc}, \ref{affil:cua}}
  \and M.~B\"ock\inst{\ref{affil:mpifr}}
  \and F.~Krau\ss\inst{\ref{affil:remeis},\ref{affil:wuerzburg}}
  \and G.~B.~Taylor\inst{\ref{affil:unm}}
  \and J.~Wilms\inst{\ref{affil:remeis}}
  \and J.~Blanchard\inst{\ref{affil:chile}}
  \and B.~Carpenter\inst{\ref{affil:cua}}
  \and T.~Dauser\inst{\ref{affil:remeis}}
  \and M.~Dutka\inst{\ref{affil:cua}}
  \and P.~G.~Edwards\inst{\ref{affil:csiro}}
  \and N.~Gehrels\inst{\ref{affil:nasa_gfsc}}
  \and C.~Gro\ss berger\inst{\ref{affil:remeis},\ref{affil:wuerzburg}}
  \and H.~Hase\inst{\ref{affil:bkg}}
  \and S.~Horiuchi\inst{\ref{affil:csiro_canberra}}
  \and A.~Kreikenbohm\inst{\ref{affil:wuerzburg},\ref{affil:remeis}}
  \and J.~E.~J.~Lovell\inst{\ref{affil:tasmania}}
  \and W.~McConville\inst{\ref{affil:nasa_gfsc}}
  \and C.~Phillips\inst{\ref{affil:csiro}}
  \and C.~Pl\"otz\inst{\ref{affil:bkg}}
  \and T.~Pursimo\inst{\ref{affil:not}}
  \and J.~Quick\inst{\ref{affil:hartrao}}
  \and E.~Ros\inst{\ref{affil:valencia1},\ref{affil:valencia2},\ref{affil:mpifr}}
  \and R.~Schulz\inst{\ref{affil:remeis},\ref{affil:wuerzburg}}
  \and J.~Stevens\inst{\ref{affil:csiro}}
  \and S.~J.~Tingay\inst{\ref{affil:curtin}}
  \and J.~Tr\"ustedt\inst{\ref{affil:wuerzburg},\ref{affil:remeis}}
  \and A.K.~Tzioumis\inst{\ref{affil:csiro}}
   \and J.~A.~Zensus\inst{\ref{affil:mpifr}}
  }

\institute{
  Dr. Remeis Sternwarte \& ECAP, Universit\"at Erlangen-N\"urnberg, 
  Sternwartstrasse 7, 96049 Bamberg, Germany \label{affil:remeis}
  \and
  Institut f\"ur Theoretische Physik und Astrophysik, Universit\"at
  W\"urzburg, Am Hubland, 97074 W\"urzburg, Germany
  \email{cornelia.mueller@sternwarte.uni-erlangen.de} \label{affil:wuerzburg}
  \and
  NASA, Goddard Space Flight Center, Greenbelt, MD 20771, USA  \label{affil:nasa_gfsc}
  \and 
  Catholic University of America, Washington, DC 20064, USA \label{affil:cua}
    \and
  Max-Planck-Institut f\"ur Radioastronomie, Auf dem H\"ugel 69, 53121 Bonn, Germany
  \label{affil:mpifr}
    \and
  Department of Physics and Astronomy, University of New Mexico,
  Albuquerque, NM 87131, USA \label{affil:unm}
  \and
  Departamento de Astronomía Universidad de Concepci\'on, Casilla 160 C, Concepci\'on, Chile
  \label{affil:chile}
    \and
  CSIRO Astronomy and Space Science, ATNF, PO Box 76 Epping, NSW 1710, Australia \label{affil:csiro}
  \and
  Bundesamt f\"ur Kartographie und Geod\"asie, 93444 Bad K\"otzting, Germany \label{affil:bkg}
  \and
  CSIRO Astronomy and Space Science, Canberra Deep Space
  Communications Complex, P.O. Box 1035, Tuggeranong, ACT 2901,
  Australia \label{affil:csiro_canberra}
\and 
  School of Mathematics \& Physics, University of Tasmania, Private Bag 37, Hobart, Tasmania 7001, Australia \label{affil:tasmania} 
  \and
  Nordic Optical Telescope Apartado 474, 38700 Santa Cruz de La Palma
  Santa Cruz de Tenerife, Spain \label{affil:not}
  \and
  Hartebeesthoek Radio Astronomy Observatory, Krugersdorp, South
  Africa \label{affil:hartrao}
  \and
  Observatori Astron\`omic, Universitat de Val\`encia, Parc Cient\'{\i}fic, C.\ Catedr\'atico Jos\'e Beltr\'an 2, 46980 Paterna, Val\`encia, Spain \label{affil:valencia1}
  \and
  Departament d'Astronomia i Astrof\'{\i}sica, Universitat de Val\`encia, C.\ Dr.\ Moliner 50, 46100 Burjassot, Val\`encia, Spain \label{affil:valencia2}
  \and
  International Centre for Radio Astronomy Research, Curtin University, 6102 Perth, Australia \label{affil:curtin}
  }

\date{draft of \today}

  \nolinenumbers
  \abstract
  {
    We investigate the nature and classification of \pmn, a bright
    radio source close to the Galactic plane, which is associated with
    one of the brightest hard-spectrum $\gamma$-ray sources detected
    by \textsl{Fermi}/LAT. It has previously been classified as a 
      low-peaked BL\,Lac object based on its broadband emission and
      the absence of optical emission lines. Optical measurements,
      however, suffer strongly from extinction and the absence of
      pronounced short-time $\gamma$-ray variability over years of
      monitoring is unusual for a blazar.
    }
    {
  In this paper, we are combining new and archival multiwavelength data of 
  \pmn in order to reconsider the classification and nature of this unusual $\gamma$-ray source.}
    { 
      For the first time, we study the radio morphology of \pmn at
      8.4\,GHz and 22.3\,GHz, and its spectral properties on
      milliarcsecond scales, based on VLBI observations from the
      TANAMI program. We combine the resulting images with
      multiwavelength data in the radio, IR, optical/UV, X-ray, and
      $\gamma$-ray regimes. }
  {
    \pmn shows a symmetric brightness distribution at 8.4\,GHz on
    milliarcsecond scales, with the brightest, and most compact component
    in the center of the emission region. The morphology is
    reminiscent of a Compact Symmetric Object (CSO). Such objects,
    thought to be young radio galaxies, have been predicted to produce
    $\gamma$-ray emission but have not been detected as a class by the
    \textsl{Fermi} $\gamma$-ray telescope so far. Sparse
    $(u,v)$-coverage at 22.3\,GHz prevents an unambiguous modeling of
    the source morphology at this higher frequency.
    Moreover, infrared measurements reveal an excess in
    the spectral energy distribution (SED), which
    can be modeled with a blackbody with a temperature of about 
    1600\,K, and which is usually not present in blazar SEDs.
    }
  {
    The TANAMI VLBI data and the shape of the broadband SED challenge
    the current blazar classification of one of the brightest
    $\gamma$-ray sources in the sky. \pmn seems to be either a highly
    peculiar BL\,Lac object or a misaligned jet source. In the latter
    case, the intriguing VLBI structure opens room for a possible
    classification of \pmn as a $\gamma$-ray bright CSO.
    }
  
  \keywords{Galaxies: active -- Galaxies: individual: PMN\,J1603$-$4904 -- Radio Continuum: general -- Techniques: interferometric -- Gamma rays: galaxies -- X-rays: galaxies}

\authorrunning{C.\,M\"uller et al.}
\titlerunning{The unusual multiwavelength properties of PMN\,J1603$-$4904}
   \maketitle


\section{Introduction}\label{sec:intro}

Blazars (flat spectrum radio quasars and BL\,Lac objects) are by far the largest class of
\textsl{Fermi}/LAT detected extragalactic $\gamma$-ray sources
\citep{2fgl}. They are a subclass of active galactic nuclei (AGN),
where the angle between the line of sight and the relativistic jet is
small. A more elusive class of $\gamma$-ray sources are extragalactic
jets seen side-on where relativistic beaming effects are small
\citep[][]{Abdo2010_misaligned}. One sub-class of such misaligned
objects are young radio galaxies, also referred to as Compact
Symmetric Objects \citep[CSO;][]{Odea1998, Readhead1996b}.
CSOs are not yet confirmed as $\gamma$-ray loud, however, theoretical
models have predicted $\gamma$-ray emission from these objects
\citep{Stawarz2008, Kino2007, Kino2009, Kino2011} with expected
$\gamma$-ray variability time scales substantially longer than that
typically observed from blazars \citep{Abdo2010_LC_LBAS}.

The radio source \object{PMN\,J1603$-$4904}
\citep[][\object{PKS\,B\,1600$-$489}]{PMN1994} has been classified as
a low synchrotron peaked (LSP) BL\,Lac object \citep{2fgl}. Located
close to the Galactic plane ($l = 332\fdg15$, $b = 2\fdg57$), no
redshift measurement has been reported to date \citep{Shaw2013a}. It
is associated with a bright, hard-spectrum $\gamma$-ray source
detected by \textsl{Fermi}/LAT and called 0FGL\,J1604.0$-$4904,
1FGL\,J1603.8$-$4903, 2FGL\,J1603.8$-$4904, and \object{1FHL
  J1603.7$-$4903}, respectively
\citep{0fgl,1fgl,2fgl,2lac,1fhl}. The radio source lies well
within the \textsl{Fermi}/LAT 95\,\%-error radius of only $0\fdg023$
\citep{2fgl} and no further known radio source lies in the vicinity,
making this a very high confidence association. Among the 30 brightest
objects in the 2LAC-catalog \citep{2lac}, PMN\,J1603$-$4904 is one out
of only two sources classified as non-variable (between 1\,GeV and
100\,GeV) with a flux of
$F_\mathrm{1-100\,GeV}=1.29\times10^{-8}\,\mathrm{ph}\,\mathrm{cm}^{-2}\,\mathrm{s}^{-1}$
over the first two years of \textsl{Fermi}/LAT observations. It was
reported as non-variable with a $\mathrm{TS}_\mathrm{var}=39.254$,
where $\mathrm{TS}_\mathrm{var}>41.64$ indicates a $>99$\%-chance of
variability. The second non-variable source is 4C\,+55.17. This source
has $\gamma$-ray properties similar to \pmn. See
\citet{McConville2011} for a discussion of the emission properties of
4C\,+55.17 in the context of a CSO model. At higher $\gamma$-ray
energies, \pmn shows a hard spectral index of
\mbox{$\Gamma_{10-500\,\mathrm{GeV}}=1.96\pm 0.14$} and a high flux
above 50\,GeV of \mbox{$S_{>50\,\mathrm{GeV}}=(19\pm
  5)\times10^{-11}\,\mathrm{ph}\,\mathrm{cm}^{-2}\,\mathrm{s}^{-1}$}
\citep{1fhl}. A Bayesian block analysis of the 10\,GeV to 500\,GeV
emission over three years of \textsl{Fermi}/LAT monitoring shows
evidence for mild variability on longer timescales but no flaring on
timescales of days to weeks \citep{1fhl, 2fgl}.

PMN\,J1603$-$4904 was observed in several large radio surveys
\citep{PMNsurvey1993, PMNsurvey1994, Massardi2011}, but no high
resolution observations with Very Long Baseline Interferometry (VLBI)
have been reported previously. We added the source to the
TANAMI\footnote{Tracking Active Galactic Nuclei with Austral
  Milliarcsecond Interferometry,
  \url{http://pulsar.sternwarte.uni-erlangen.de/tanami}} sample in
2009 \citep{Ojha2010a, Mueller2012a}. TANAMI is a large VLBI program,
currently monitoring 84 extragalactic jets south of $-30^\circ$
declination at 8.4\,GHz and 22.3\,GHz. The initial TANAMI sample was
defined combining a radio and $\gamma$-ray selected sub-sample. Since
the launch of \textsl{Fermi}, newly detected $\gamma$-ray and
radio-bright extragalactic radio sources have been included in the
program. For most of these sources TANAMI provides the first-ever VLBI
images at milliarcsecond (mas) resolution \citep{Mueller2012a,
  Mueller2013a}.

In Sect.~\ref{sec:obs}, we describe the TANAMI VLBI observations and
report on the analysis of additional radio monitoring and
\textsl{Swift}/XRT data. Then we present the results derived from
first images of the milliarcsecond-scale structure of PMN\,J1603$-$4904 and its
broadband spectral energy distribution (SED). In
Sect.~\ref{sec:discussion}, we discuss the VLBI morphology and the
multiwavelength properties, which are difficult to reconcile within a
blazar classification scenario. We therefore also discuss alternative
scenarios, including a classification as a young radio galaxy.

%
\section{Observations}\label{sec:obs}
\subsection{TANAMI VLBI Observations}

As part of the extended TANAMI sample \citep{Mueller2012a,
  Boeck2012PhDT}, PMN\,J1603$-$4904 has been observed for three times
between 2009 February and 2010 May at 8.4\,GHz, including one
simultaneous dual-frequency observation at 8.4\,GHz and 22.3\,GHz in
2010 May. TANAMI is using the Australian Long Baseline Array (LBA) and
additional radio telescopes in Antarctica, Chile, New Zealand and
South Africa \citep{Ojha2010a}. Details of the observations are given
in Table~\ref{table:1}. Data were correlated on the DiFX software
correlator \citep{Deller2007, Deller2011} at Curtin University in
Perth, Western Australia, and calibrated and imaged as described in
\citet{Ojha2010a}.

\begin{figure}
    \includegraphics[width=\columnwidth]{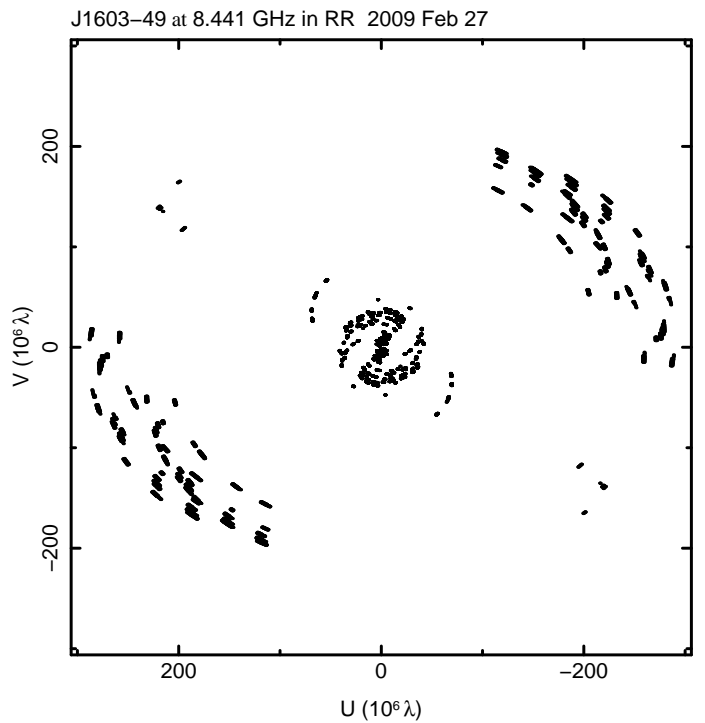}
    \caption{$(u,v)$-coverage at 8.4\,GHz of the combined 2009 February TANAMI
      observations (see Table~\ref{table:1} for
      details).\label{fig:uvplot}}
\end{figure}

Because of the different number of non-LBA telescopes in the different
experiments, the $(u,v)$-coverage, array sensitivity, and angular
resolution vary between the observing epochs. Figure~\ref{fig:uvplot}
shows the $(u,v)$-coverage for the 2009 February observation at
8.4\,GHz, which involved the largest number of antennas in all
observations.

\begin{table*}
    \caption{8.4 and 22.3\,GHz TANAMI VLBI observations of
      PMN\,J1603$-$4904 and image parameters} 
    \label{table:1}      
    \centering          
      \resizebox{\textwidth}{!} {
    \begin{tabular}{cclcccccc}    
        \hline\hline 
        Date &  Freq. & Array Configuration\tablefootmark{a} & $S_\mathrm{peak}$ & RMS & $S_\mathrm{total}$ & $b_\mathrm{maj}$\tablefootmark{c} & $b_\mathrm{min}$\tablefootmark{c} & P.A.\tablefootmark{c} \\
        yyyy-mm-dd  & [GHz]&            & [Jy beam$^{-1}$] &  [mJy beam$^{-1}$] & [Jy] &  [mas] & [mas] & [$^\circ$] \\ 
        \hline    
        2009-02-23/27\tablefootmark{b} & 8.4 & PKS-CD-HO-MP-AT-DSS43-DSS34-TC-OH & 0.18  & 0.13   & 0.59 &    2.51 & 0.98 &  30.1\\
        2009-09-06 & 8.4 &PKS-CD-HO-MP-DSS43-TC         & 0.17  & 0.23  &0.57  & 3.36 & 1.15    &  19.2 \\
        2010-05-07 & 8.4 &PKS-CD-HO-MP-AT-TC            & 0.17  & 0.16 & 0.57    &  2.91 & 1.12&  14.9\\
        \hline
        
        2010-05-05  & 22.3 &PKS-HO-MP-AT                & 0.14  & 0.35 & 0.29   &2.09  &1.16    & 83.9  \\
        \hline \hline 
    \end{tabular}
    }
    \tablefoot{ \tablefoottext{a}{AT: Australia Telescope Compact
      Array (ATCA), CD: Ceduna, HO: Hobart, MP: Mopra, OH:
      GARS/O'Higgins, PKS: Parkes, TC: TIGO, DSS34\,\&\,43:
      Tidbinbilla (34\,m \& 70\,m)} 
     \tablefoottext{b}{The two consecutive 2009 February experiments
       were combined, due to poorer $(u,v)$-sampling compared to the
       subsequent experiments at 8\,GHz.} 
   \tablefoottext{c}{Major, minor axis and position angle of the
     restoring beam.} 
      }
\end{table*}

\subsection{ATCA radio observations}
The TANAMI sample is regularly monitored with the Australia Telescope
Compact Array (ATCA). ``Snapshot'' observations are made at 5.5, 9,
17, 19, 38 and 40\,GHz, where each frequency is the center of a 2\,GHz
wide band and the flux densities are calibrated against the ATCA
primary flux calibrator PKS\,1934$-$638 and/or Uranus at 7\,mm
\citep{Stevens2012}. The primary beam widths of the ATCA radio
telescope at the observed frequencies range from $10'$ at 5.5\,GHz to
$1'$ at 40\,GHz. The ATCA flux densities are found to be independent
of the exact array configuration, indicating a compact nature for
PMN\,J1603$-$4904 at kiloparsec scales.

\subsection{Optical, UV, and X-ray observations}\label{susec:mwobs}
\pmn has been observed three times by the \textsl{Swift} satellite
\citep{Gehrels2004} between 2009 May and 2010 July for a total
exposure of 6.15\,ks. The \textsl{Swift} UV/Optical Telescope
\citep[UVOT,][]{Roming2005} observed the source at optical/ultraviolet
wavelengths with two filters (UVM2 and U). The source could not be
detected at the $3\sigma$ level and we only report upper limits on the
flux, which we estimated from the background. Simultaneous to the UVOT
observations, \textsl{Swift} X-ray Telescope
\citep[XRT,][]{Burrows2005} observations reveal an X-ray source which
is positionally consistent with the radio source. No other archival
X-ray observations are available. The \textsl{Swift}/XRT data were
processed using the XRTDAS software package, which is part of the
HEASoft package (v6.13). We reduced the data using the
\texttt{xrtpipeline} task. The extraction of the data was performed by
selecting a $47\farcs15$ extraction radius centered on the source
coordinates and an annulus for the background region with radii of
$80''$ and $120''$. 
We used the accumulated data of all three
observations\footnote{A separate X-ray data analysis of each single \textsl{Swift}/XRT
observation was not possible.}, 
which yielded a total of 77 source photons. Due to this
small number statistics, we assume a standard\footnote{for our TANAMI sample \citep[compare also][]{Kadler2005}} 
absorbed power law ($\Gamma=1.64$) and an absorption of
$N_\mathrm{H}=6.32\times10^{21}\,\mathrm{cm}^{-2}$ according to the Galactic H\textsc{i} value 
in this direction \citep{Kalberla2005}. 
The observed X-ray counts yield then a roughly estimated flux of $\sim5\times
10^{-13}\,\mathrm{erg}\,\mathrm{cm}^{-2}\,\mathrm{s}^{-1}$ in the
0.5--10\,keV band.

\pmn was observed with the Gemini South telescope using the Gemini
Multi-Object Spectrographs (GMOS) on 2013 March 10. The observation
was made in snapshot mode in the r'\,band (centered at $\lambda\sim
6250$\AA) to identify the optical counterpart of the radio source and
to facilitate spectroscopic followup. The source was observed for 120
seconds, under $0\farcs6$ seeing and a limiting magnitude of about
25\,mag was achieved.


\section{Results}\label{sec:results}

\subsection{Brightness distribution on milliarcsecond scales}
\begin{figure}
    \includegraphics[width=\columnwidth]{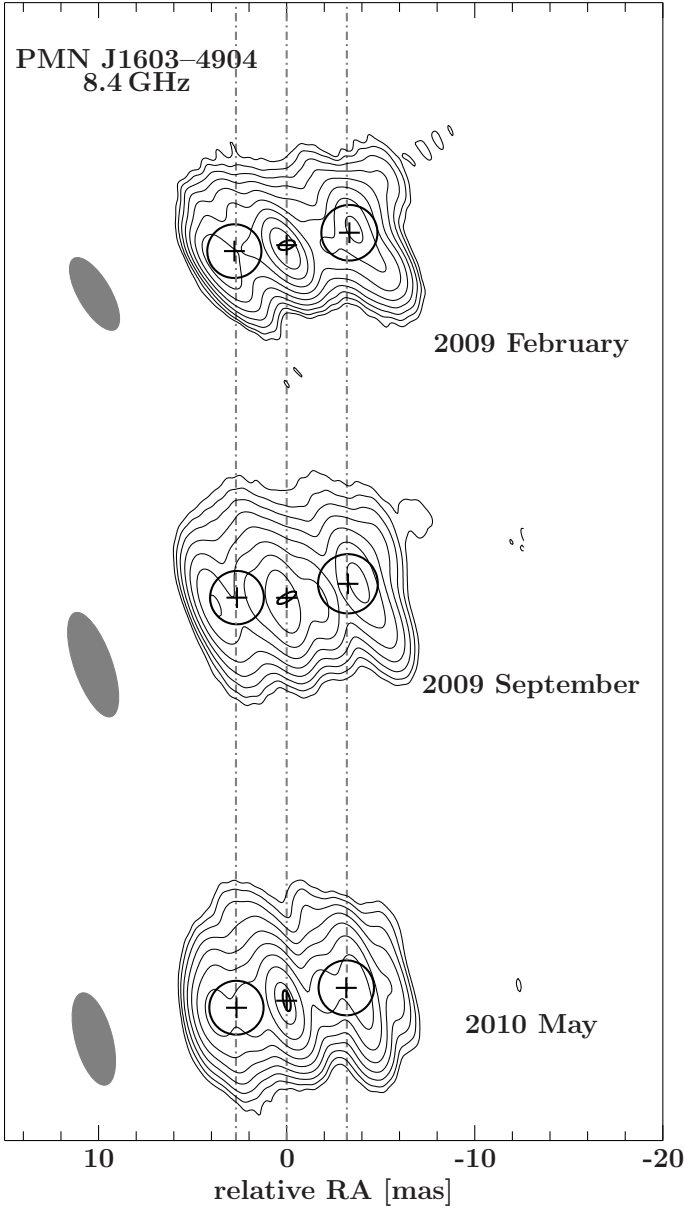}
    \caption{Time evolution of PMN\,J1603$-$4904. CLEAN images of the
      first 8.4\,GHz TANAMI observations are shown. The contours
      indicate the flux density level, scaled logarithmically and
      separated by a factor of 2, with the lowest level set to the
      3$\sigma$-noise-level (for more details see
      Table~\ref{table:1}). The positions and FWHMs of the Gaussian
      emission components are overlaid as black ellipses (for model
      parameters see Table~\ref{table:2}). From top to bottom: 2009
      February (combined image of $23^\mathrm{th}$ and
      $27^\mathrm{th}$), 2009 September, and 2010 May. The size of the
      restoring beams for each individual observation is shown as gray
      ellipse in the lower left corners. Vertical dashed lines which
      indicate the relative positions of the eastern and western
      features with respect to the central component are drawn at of
      2.7\,mas and $-3.2$\,mas, respectively. \label{fig:allepochs} }
\end{figure}

Table~\ref{table:1} shows the results of hybrid imaging using the
\texttt{CLEAN} algorithm \citep{Hoegbom1974}. The three 8.4\,GHz VLBI
observations of PMN\,J1603$-$4904 show an almost constant total
correlated flux density of $570$\,mJy to $590$\,mJy. Changes in the
measured values are well below the (conservative) estimate of absolute
calibration uncertainties of $\sim$20\% \citep{Ojha2010a}. The images
have a dynamic range (ratio of peak brightness to five times the
root-mean-squared noise level) of $\sim$80 to $\sim$280.

The source is resolved in all directions and shows an east-west
orientation with three resolved distinct regions within
$\lesssim$15\,mas. The brightest, most compact feature is located in
the center (see Fig.~\ref{fig:allepochs}).

In order to study the flux density and structural variability of the
individual prominent jet features, we choose a simple model of three
elliptical Gaussian emission components \citep[using
\texttt{DIFMAP},][]{Shepherd1997} to fit the self-calibrated
visibility data. The parameters of the model components are reported
in Table~\ref{table:2}. The central component is very small but
resolved \citep[see][for details on the resolution
limit]{Kovalev2005}. Using
\begin{equation}
    T_\mathrm{B} = \frac{2\ln 2}{\pi k}\frac{S \lambda^2}{a_\mathrm{maj}a_\mathrm{min}} \cdot (1+z)
\end{equation}
where $S$ is the flux density, $a_\mathrm{maj, min}$ are the
major and minor axes of the component, $k$ is the Boltzmann constant, $z$ is the redshift, and $\lambda$ is
the wavelength of the observation \citep{Kovalev2005}, the brightness
temperature of this brightest of all found components is
$T_{\mathrm{B, central}} \gtrsim 9 \times 10^{9}$\,K (due to the lack of a redshift measurement, we apply $z=0$, i.e., no redshift correction). No component
shows significant flux density or brightness temperature variability
over 15\,months. The eastern and western regions can each be modeled
with circular Gaussian flux distributions. The eastern component is
about $(0.2\pm 0.1)$\,Jy weaker and about 0.6\,mas closer to the
central component than the western one. The brightness temperatures of
both outer components are also constant at $\mbox{(3--4)} \times
10^{8}$\,K.

To test for structural variability, we measured component positions
relative to the eastern component. Within the uncertainties, no
significant component motions could be found over the covered period
of 15 months. Therefore we can set a limit for the relative motions of
$v_\mathrm{app}<0.2\,\mathrm{mas}\,\mathrm{yr}^{-1}$.


\begin{table}
    \caption{Modelfit parameters for TANAMI VLBI images\label{table:2} }
    
    \renewcommand{\arraystretch}{1.1}
    \textbf{8.4\,GHz}\\
    \begin{tabular}{lllllll}
        \hline
        \hline
        $S$\tablefootmark{a} & $d$\tablefootmark{b}  & $\theta$\tablefootmark{b} &
        $a_\mathrm{maj}$\tablefootmark{c}  & $a_\mathrm{min}$\tablefootmark{c} &
        P.A.\tablefootmark{c}  & $T_B$\tablefootmark{d}  \\
        $[\mathrm{Jy}]$& [mas] & [$^\circ$] & [mas] &  [mas] &
        [$^\circ$] & [$10^9$\,K] \\
        \hline%
        \multicolumn{7}{l}{ Combined 2009-02-23 \& 2009-02-27:}\\
        \hline
        0.17 & 2.8 & 97.0 & 2.9 & 2.9 & 138.0 & 0.35 \\
        0.22 & 0.0 & --- & 0.9 & 0.5 & 33.0 & 9.1 \\
        0.20 & 3.4 & $-$79.0 & 3.0 & 3.0 & $-$114.0 & 0.39 \\
        \hline
         \multicolumn{7}{l}{2009-09-05:}\\
        \hline
        0.16 & 2.6 & 90.0 & 2.9 & 2.9 & $-$155.0 & 0.34 \\
        0.22 & 0.0 & --- & 1.1 & 0.4 & 17.0 & 8.6 \\
        0.19 & 3.3 & $-$77.0 & 3.2 & 3.2 & $-$153.0 & 0.32 \\
        \hline
         \multicolumn{7}{l}{2010-05-07:}\\
        \hline
        0.17 & 2.7 & 98.0 & 3.0 & 3.0 & 177.0 & 0.35 \\
        0.21 & 0.0 & --- & 1.1 & 0.4 & 12.0 & 8.6 \\
        0.19 & 3.2 & $-$78.0 & 2.9 & 2.9 & 20.0 & 0.38 \\
        \hline
    \end{tabular}
    
    \vspace{0.3cm}
    \textbf{22.3\,GHz}\\
    \begin{tabular}{lllllll}
        \hline
        \hline
        $S$\tablefootmark{a} & $d$\tablefootmark{b}  & $\theta$\tablefootmark{b} &
        $a_\mathrm{maj}$\tablefootmark{c}  & $a_\mathrm{min}$\tablefootmark{c} &
        P.A.\tablefootmark{c}  & $T_B$\tablefootmark{d}  \\
        $[\mathrm{Jy}]$& [mas] & [$^\circ$] & [mas] &  [mas] &
        [$^\circ$] & [$10^9$\,K] \\
        \hline
        \multicolumn{7}{l}{2010-05-05:}\\
        \hline
        0.04 & 2.7 & 98.0 & 3.0 & 3.0 & 177.0 & 0.01 \\
        0.19 & 0.0 & --- & 1.1 & 0.4 & 12.0 & 1.1 \\
        0.07 & 3.2 & $-$78.0 & 2.9 & 2.9 & 20.0 & 0.02 \\
        \hline
    \end{tabular}
\tablefoot{ 
      \tablefoottext{a}{Integrated flux density of model component.} 
     \tablefoottext{b}{Distance and position angle of the model component from the designated
      phase center}
     \tablefoottext{c}{Major, minor axis extent (FWHM) of the model component and position angle of the major axis.}
      \tablefoottext{d}{Brightness temperature of model component.}
  }
\end{table}

\subsection{Spectral properties on mas-scale}\label{susec:spix}
In 2010 May, contemporaneous 8.4\,GHz and 22.3\,GHz TANAMI VLBI
observations were performed (see Table~\ref{table:1}). The
$(u,v)$-coverage at 22.3\,GHz is poorer than that at 8.4\,GHz, because
the TIGO antenna is not equipped with a 22.3\,GHz receiver and the
Ceduna data were not usable due to problems with the maser, such that
effectively only data from four Australian antennas were available. In
order to image and self-calibrate the 22.3\,GHz $(u,v)$-data, we used
the structural model found from the high-quality data at 8.4\,GHz. We
fixed the relative positions of the three components from the 8.4\,GHz
Gaussian model and allowed only their flux densities to vary. This
approach resulted in an acceptable starting model to perform amplitude
self-calibration on long time scales. We then fitted the component
flux densities again and performed self-calibration on iteratively
smaller time scales. Overall self-calibration corrections were small
and the final model was in good agreement with the original data. 
This
model represents the most extended structure, which is still
consistent with both the $(u,v)$-data at 22.3\,GHz itself and with the
brightness distribution found at 8.4\,GHz. More extended regions in
VLBI images of extragalactic jets have steep spectra so the
22.3\,GHz emission region is unlikely to be larger than the 8.4\,GHz emission
region. For these reasons, we refer to this model of the 22.3\,GHz
brightness distribution as the `extended' model.

\begin{figure}
    \includegraphics[width=\columnwidth]{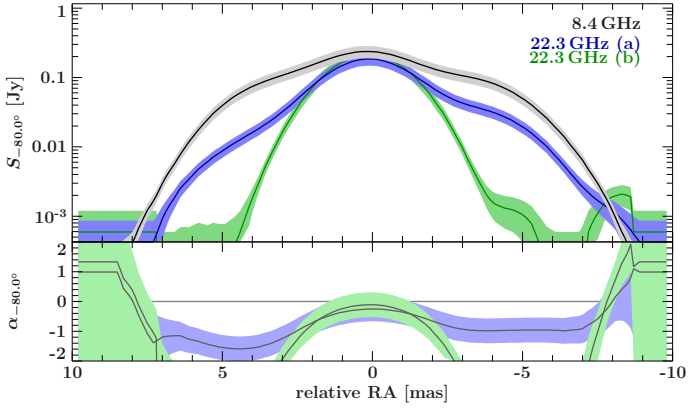}
    \caption{Top: Flux density profiles along P.A.$=-80^\circ$ at
      8.4\,GHz (gray) and the 22.3\,GHz `extended model' (blue, (a))
      and `compact model' (green, (b)). Bottom: Spectral index along
      P.A.$=-80^\circ$. Displayed uncertainties correspond to a conservative
      estimate of absolute calibration uncertainties and on-source
      errors of $\sim$20\%. The spectral index distribution of 
      the `extended model' in light blue, the `compact
      model' is shown in light green.}\label{fig:spixcut}
\end{figure}

In order to estimate the spatial spectral index\footnote{The spectral
  index $\alpha$ is defined through $S_\nu\propto\nu^{+\alpha}$.}
distribution, we combined the 8.4\,GHz image of 2010 May with the
quasi-simultaneous 22.3\,GHz image, which was produced from the
`extended model' (Fig.~\ref{fig:spixcut}). We convolved both data sets
with a common circular beam with a major axis of 3\,mas and calculated
cuts through the two brightness distributions along
$\mathrm{P.A.}=-80^\circ$. Formally, we can assign uncertainties to
both profiles according to the absolute calibration uncertainties in
both frequency bands. This approach, however, would not account for
the much larger structural uncertainties at 22.3\,GHz because of the
sparse $(u,v)$-coverage. To constrain these uncertainties, we derived
a `compact' model of the 22.3\,GHz brightness distribution by
considering only one elliptical Gaussian component in the
model-fitting/self-calibration cycle described above. The resulting
profile of the `compact' model is also shown in
Fig.~\ref{fig:spixcut}.

The bottom panel of Fig.~\ref{fig:spixcut} shows the formal spectral
index profile calculated for the `extended' and `compact' models. It
is obvious that the sparse $(u,v)$-coverage affects mainly the eastern
and western wings of the spectral index profile. The central component
is consistently found at a spectral index of $-0.75 \lesssim \alpha
\lesssim -0.25$. The eastern and western regions have steep spectra in
the `extended model' with $-2.0 \lesssim \alpha \lesssim -1.0$, but
reach unrealistically steep values in the `compact model' ($\alpha \ll
-2.0$). We conclude that the compact model, while formally
representing the sparse $(u,v)$-data well, does not lead to a
self-consistent dual-frequency model.

Because absolute position information is lost in the self-calibration
process, the `compact model' would in principle allow different
alignments with the 8.4\,GHz image. In particular, in a blazar
scenario (see Sect.~\ref{sec:discussion}), one might consider the peak
of the 22.3\,GHz emission in the `compact model' to be associated with
the eastern source component at 8.4\,GHz (i.e., a shift of
$\sim$2.7\,mas) since the tapered 8.4\,GHz image shows a potential,
faint extension to the west. With this alternate alignment, the
putative blazar core in the east would have a slightly inverted
spectrum, but the bright 8.4\,GHz emission westward would reach again
unphysical, extremely steep spectral-index values ($\alpha \ll -4$).

These considerations reveal the central feature as the most plausible
`core' of \pmn, as the region with the flattest spectral index and the
highest brightness temperature.

\subsection{Integrated ATCA radio spectral monitoring}\label{susec:atca}

\begin{figure}
    \includegraphics[width=\columnwidth]{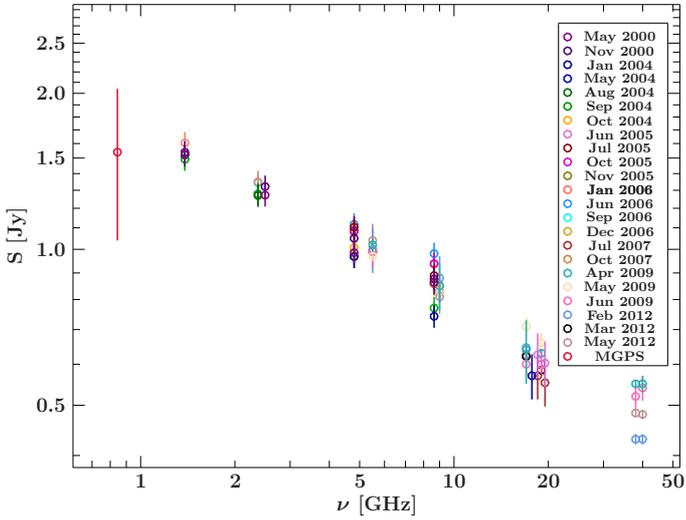}
    \caption{Radio spectrum of \pmn at 6 frequencies based on the ATCA
      monitoring. The error bars are calculated as a
      frequency-dependent fraction of the flux density at each
      frequency. The archival data of the MGPS catalog might indicate
      a turnover  of the spectrum below 0.8\,GHz.}\label{fig:C1730SPEC}
\end{figure}

\begin{figure}
    \includegraphics[width=\columnwidth]{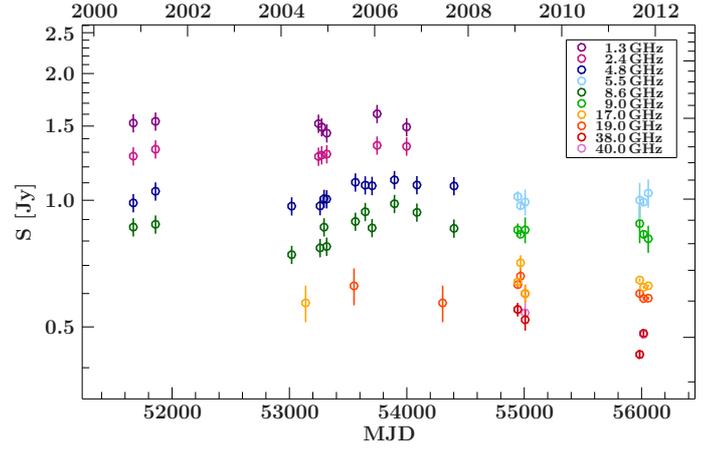}
    \caption{Light curve of \pmn from the ATCA monitoring at 6
      frequencies. Error bars as in Fig.~\ref{fig:C1730SPEC}. Note the
      lack of variability between 2009 and 2012. On longer timescales
      some evidence for low-level variability is seen.}\label{fig:C1730LC}
\end{figure}

In Fig.~\ref{fig:C1730SPEC} and \ref{fig:C1730LC} we show the radio
spectra and light curves which were collected as part of the C1730
ATCA program \citep{Stevens2012} as well as of archival ATCA
observations. Between 2009 and 2012 the source shows no significant
variability. Its spectrum has a spectral index of $\alpha \sim -0.4$,
consistent with the one determined from the integrated VLBI images
($-1.0 \lesssim \alpha \lesssim 0.0$, assuming the `extended model' at
22.3\,GHz). Note, however, that the ATCA flux densities are
$\sim$200\,mJy brighter. This indicates a diffuse extended emission component 
which is resolved out by the TANAMI VLBI array.

\subsection{Multiwavelength Source Association}\label{susec:confusion}
The association of the \textsl{Fermi}/LAT source 2FGL\,J1603.8$-$4904
with the radio object \pmn is well established \citep{0fgl,1fgl,2fgl}.
The Molonglo Galactic Plane Survey catalog lists the source with
\mbox{$S_\mathrm{843\,MHz} =(1.54\pm0.5)$\,Jy} \citep[MGPS, source ID:
J160350$-$490403,][]{Murphy2007_MGPS}. \textsl{Swift}/XRT observations
reveal a single, but faint, X-ray source. While the possibility of
association with a spurious source within the \textsl{Swift}/XRT error
circle cannot be ruled out, the radio position of \pmn clearly falls
within the \textsl{Swift}/XRT's point spread function (PSF) of $18''$
Half Power Diameter (HPD) at 1.5\,keV. Combined with the favorable
extrapolation of the nonthermal X-ray emission into the $\gamma$-ray
regime (see Sect.~\ref{susec:SED} below for more details), this
positional coincidence leads us to believe that \pmn is the most
likely association of the X-ray source.

To identify the optical/IR counterpart of \pmn, we use data from
various sky surveys. In 2010, NASA's Wide-field Infrared Survey
Explorer \citep[\textsl{WISE};][]{Wright2010_WISE} mapped the sky at
3.4, 4.6, 12, and 22\,$\mu$m (bands W1, W2, W3, and W4) with
$6''$--$12''$ resolution. 
The \textsl{WISE} catalogs
flags the associated source  \object{WISE J160350.68$-$490405.6} 
as ``extended'', i.e., not being
consistent with a point source. Since the closest detected
\textsl{WISE} source has a distance of $\approx 12''$ from \pmn, the
extension must be caused by a closer object that is not resolved with
\textsl{WISE}.

\begin{figure}
  \includegraphics[width=\columnwidth]{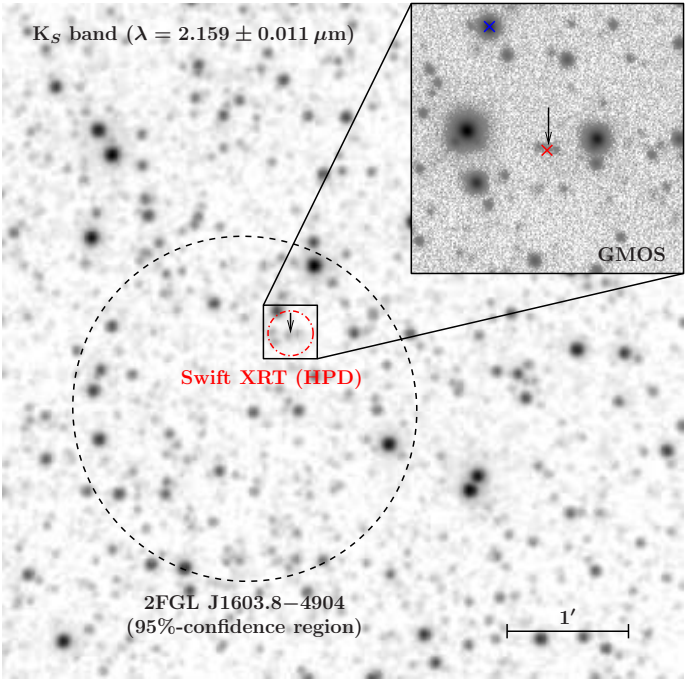}
  \caption{Grayscale 2MASS image ($\mathrm{K}_S$ band) of the region
    around the radio position of \pmn, indicated by the black arrow.
    The black-dotted circle indicates the 95\,\%-confidence ellipse of
    \textsl{Fermi}/LAT, the red-dotted circle the \textsl{Swift}/XRT
    HPD. The close-up in the north western corner shows the higher
    resolved GMOS (r'\,band) image revealing a faint
    elliptically-shaped source. The red cross marks the position of 
    the associated \textsl{WISE}/2MASS source, the blue one the
    closest source detected by \textsl{WISE} (see
    Sect.~\ref{susec:confusion}). No other known radio source is found
    in this region, indicating a high-confidence association with the
    high-energy counterpart.
    \label{fig:sky} }
\end{figure}

Figure~\ref{fig:sky} shows the higher resolved 2MASS image
\citep[][angular resolution $\sim 2''$]{Skrutskie2006_2MASS}. In this
$\mathrm{K}_S$ band (2.159\,$\mu$m) image a very faint IR source is
seen which is positionally consistent and associated with \pmn
(2MASS\,16035069$-$4904054, ID: 655163671). This counterpart is also
flagged as ``photometric confused'', likely due to a $4\farcs8$ distant
neighboring star of comparable IR magnitude. Relatively to the counterpart to \pmn, this object
becomes fainter towards longer wavelengths. This indicates that the
emission seen in \textsl{WISE} is probably dominated by \pmn, however,
in the following we will consider the infrared fluxes from WISE and
2MASS are contaminated, as we cannot exclude a contribution of this
neighboring star in these surveys (see also Sect.~\ref{susec:SED}).

Going to even higher resolution, the $r'$-band GMOS observation is
able to separate the neighboring star and \pmn. The GMOS image reveals
faint, extended emission with an East-West extension of $2\farcs5$ and
$1\farcs3$ in North-South direction at the position of \pmn. The
apparent magnitude of the Eastern part of the emission is about
24\,mag, which we will take to be the upper limit of the point source
flux density of the optical counterpart (Fig.~\ref{fig:sky}).

\citet{Shaw2013a} performed optical spectroscopic observations of all
2LAC BL\,Lac objects. \pmn was observed (on 2009 Aug 22) with the New
Technology Telescope (NTT) at La Silla Observatory. Based on low
signal-to-noise spectra these authors labeled the source as a BL\,Lac
using their standard criteria and determined a limit for the redshift
of \mbox{$z_\mathrm{max}=4.24$}. Due to its location in the Galactic plane,
these optical observations suffered from strong extinction along the
line of sight. Hence, more sensitive optical observations are required
to better constrain this limit and to obtain more information on the
optical counterpart.

Absorption is also responsible for the absence of a matching
\textsl{Swift}/UVOT counterpart (see Sect.~\ref{susec:mwobs}). With
additional positional constraints from infrared information by 2MASS
\citep{Skrutskie2006_2MASS} and \textsl{WISE} we can exclude the two
nearby UV-sources visible in the UVOT image as possible optical
counterparts of \pmn.

In the following, we assume that the reported multiwavelength
counterparts are positionally consistent and discuss on this basis the
broadband properties of \pmn.

\subsection{The broadband SED of \pmn}\label{susec:SED}
\begin{figure}
    \sidecaption
    \includegraphics[width=\columnwidth]{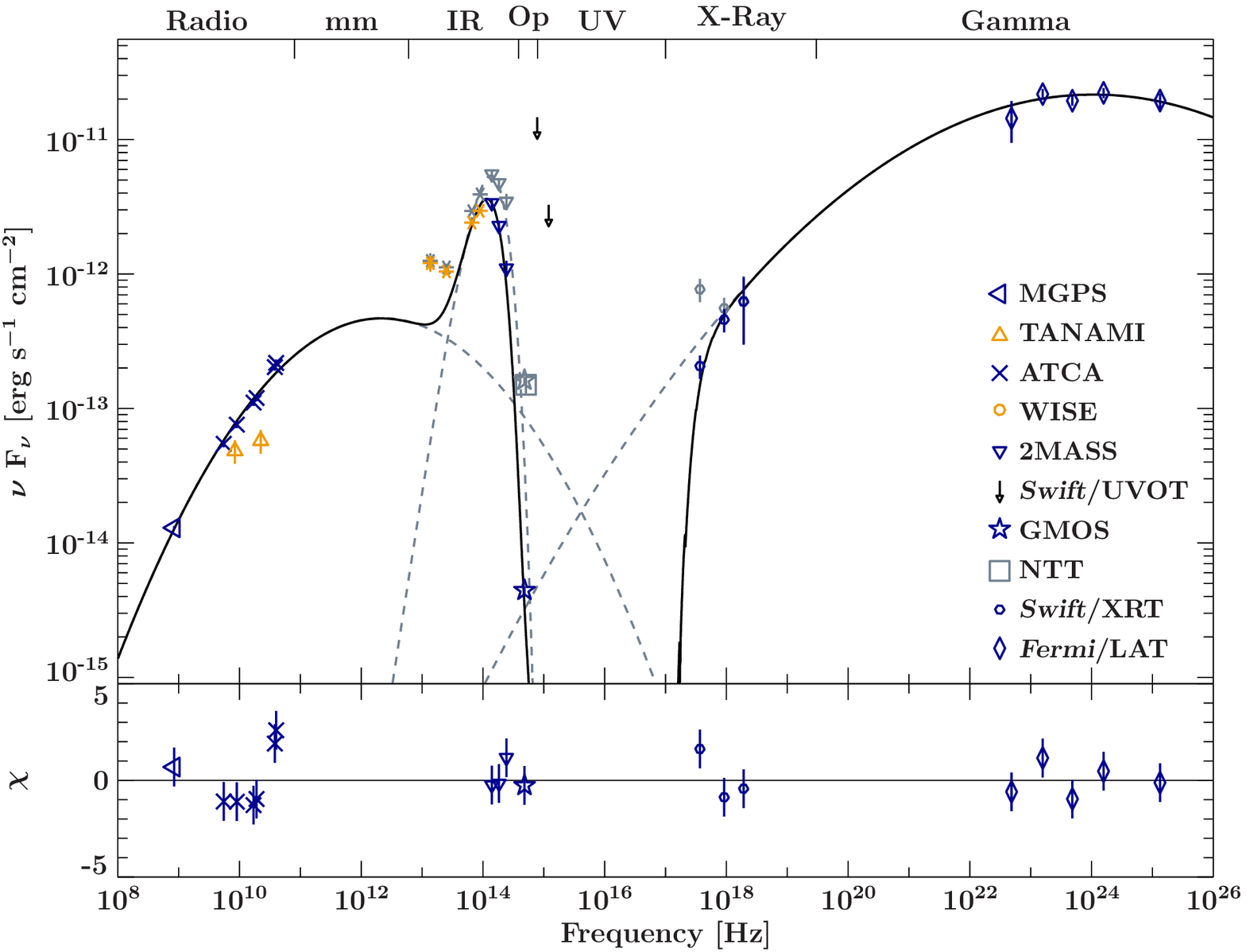}
    \caption{Broadband $\nu F_\nu$ SED including total fluxes of
      simultaneous TANAMI observations (of 2010 May), non-simultaneous
      measurements in the radio by the Molonglo Galactic Plane Survey
      \citep[MGPS,][]{Murphy2007_MGPS}, by ATCA \citep{Stevens2012},
      by \textsl{WISE} \citep{Wright2010_WISE} and 2MASS
      \citep{Skrutskie2006_2MASS}, by GMOS and NTT \citep{Shaw2013a},
      by \textsl{Swift}/XRT, UVOT, and \textsl{Fermi}/LAT
      \citep[2FGL,][]{2fgl}. 
      The data are parametrized with two logarithmic
      parabolas absorbed by photoelectric absorption at X-ray energies
      with $N_\mathrm{H} = 1.3 \times 10^{22}\,\mathrm{cm}^{-2}$ 
      and an additional blackbody component (see
      Sect.~\ref{susec:SED} for details; $\chi^2 = 23$ for 12 degrees
      of freedom).
      Blue/Orange symbols mark the data used/ignored for the SED parametrization
      (black solid line) as described in Sect.~\ref{susec:SED}.
      The IR and optical data are corrected for
      extinction \citep{Fitzpatrick1999,Nowak2012}, and the X-ray data for absorption 
      \citep{Wilms2000,Verner1996,Wilms2012},
      shown as corresponding gray symbols. 
      The gray-dashed lines represent the unabsorbed logarithmic parabolas 
      as well as the extinction corrected blackbody component.
      The bottom panel shows the residuals of the fit in
      units of the standard deviation of the individual data points.
    } \label{fig:SED}
\end{figure}

We have constructed a non-simultaneous broadband spectral energy
distribution (SED) from radio\footnote{For the SED parametrization, we used the 
ATCA rather than the VLBI data because of better frequency coverage.}, 
optical/UV, and X-ray data, as well as
published IR, optical\footnote{Catalog values corrected for extinction
  using standard values \citep[see][for more
  information]{Shaw2013a}.}, and \textit{Fermi}/LAT $\gamma$-ray data
(Fig.~\ref{fig:SED}).
Since the broadband spectral shape of \pmn resembles the typical 
double-humped SEDs of blazars, we parametrized the data with two logarithmic
parabolas \citep{Giommi2012a, Chang2010PhDT, Massaro2004}, as well as
the photoelectric absorption component at X-ray energies\footnote{Note, that the 
amount of absorption is not well constrained because of the low photon statistics in the X-ray 
band and the oversimplified continuum model. Furthermore, as the source is located in the Galactic plane, the high 
$N_\mathrm{H}$ value can be explained by ionized and/or molecular gas not measured 
by H\textsc{i} observations of the Leiden/Argentine/Bonn Survey \citep[LAB,][]{Kalberla2005}.} 
with $N_\mathrm{H} = 1.3 \times 10^{22}\,\mathrm{cm}^{-2}$
using the
\texttt{tbnew} model of \citet{Wilms2012} with abundances from
\citet{Wilms2000} and cross-sections from \citet{Verner1996}, and an
extinction model for the optical \citep{Predehl1995,Fitzpatrick1999,Nowak2012}.
This results in a synchrotron peak frequency of $\nu_\mathrm{sync}
\simeq 2.2\times10^{12}$\,Hz.   

As discussed in Sect.~\ref{susec:confusion}, the IR fluxes are likely to be
contaminated due to source confusion, however, the
contribution of the neighboring star is expected not to exceed 50\%
(see Fig.~\ref{fig:sky}). We therefore attribute the remaining excess
to \pmn. 
The most striking feature in this SED is the strong excess of
a factor of ten in flux in the IR band. We fit this IR excess with a
black body spectrum with a temperature of $T = 1662.00^{+120.0}_{-150.0}$\,K 
to the 2MASS and
GMOS data\footnote{As the NTT data were already corrected for
  extinction using a different correction factor, only the GMOS data
  were used for the parametrization in this work. The \textsl{WISE} data were
excluded in the parametrization, due to poorer angular resolution compared to 2MASS,
declining towards longer wavelenghts of the \textsl{WISE} bands
\citep[$6\farcs1$, $6\farcs4$, $6\farcs5$, $12\farcs0$ at
3.4, 4.6, 12, and 22\,$\mu$m, respectively,][]{Wright2010_WISE}.
Nevertheless, the \textsl{WISE} data match this parametrization of the SED, further
indicating that \pmn is dominating the emission in this band.} 
giving an estimate on the blackbody
temperature of \pmn and accounting for the contribution of the 
neighboring star. This can most likely be
associated with a dusty torus, the host galaxy \citep{Malmrose2011a},
or starburst activity (see Sect.~\ref{susec:alternative}).

Given the optical source morphology, it is very unlikely that the IR
excess is a foreground star: A red giant ($L = 115 L_\odot$) with this
surface temperature would need to be at a distance of 35.7\,kpc, well
outside the disk of the Galaxy, while an M-dwarf with $L = 2.21\times
10^{-5} L_\odot$ would have to be very close at 15.7\,pc. The latter
is unlikely, since the absorption along the line of sight is
constrained by the Galactic H\textsc{i} value of $N_\mathrm{H} = 6.32
\times 10^{21}\,\mathrm{cm}^{-2}$ in this direction
\citep[][]{Kalberla2005}. A nearby M-dwarf would exhibit a significantly
smaller extinction.

\section{Discussion}\label{sec:discussion}

\pmn has traditionally been classified as a low-peaked BL\,Lac object
based on its $\gamma$-ray to low-energy properties \citep{2fgl} and
the lack of emission lines \citep{Shaw2013a}. However, the source
location close to the Galactic plane makes high resolution optical
spectroscopy difficult. In the following, we first review the blazar
scenario before we discuss possible alternative source
classifications.

\subsection{\pmn as a BL\,Lac type object}\label{susec:blazar}

Within the unification scheme of radio-loud AGN
\citep{Antonucci1993,Urry1995}, BL\,Lac objects are considered to be
the beamed version of FR\,I type galaxies \citep{Fanaroff1974}, with
jets pointing close to the line of sight. The broadband SED can
locally be well-described by a power law component. Thermal emission
features in the optical and infrared are usually outshined by the
non-thermal jet emission
\citep[e.g.,][]{Abdo2010e,Chen2011,Plotkin2012}. It is only in very
few blazars that an IR excess over the underlying power law component
can be related to a hot torus \citep{Malmrose2011a}.

BL\,Lacs are in general strongly variable across the whole
electromagnetic spectrum and show a clear double-humped SED shape.
Based on the position of the synchrotron hump, sub-classifications are
made into low- (LSP), intermediate- (ISP) and high-synchrotron-peaked
(HSP) BL\,Lacs with $\nu^S_\mathrm{peak}<10^{14}$\,Hz, $10^{14} <
\nu^S_\mathrm{peak} < 10^{15}$\,Hz and $\nu^S _\mathrm{peak} >
10^{15}$\,Hz, respectively \citep[e.g.,][]{Donato2001}. High angular
resolution VLBI measurements of BL\,Lacs in general reveal highly
polarized ($\sim$10\%) jet emission
\citep{Lister2005,Hovatta2010,Linford2011a} and apparent superluminal
motion of individual jet components
\citep{Kellermann2004,Lister2009c}. Multifrequency VLBI observations
\citep[e.g.,][]{Taylor2005} of mas-scale blazar jets show that the
radio spectral index distribution of the core is flat to inverted,
i.e., self-absorbed while outer jet emission becomes optically thin.

Within the blazar-classification scenario, multiple observational
facts of \pmn point to a highly peculiar object:

\paragraph{Variability:}

Multiwavelength monitoring of \pmn reveals no short-term or rapid
flaring variability. After two years of \textsl{Fermi}/LAT monitoring,
in the 2FGL catalog \citep{2fgl}, \pmn was listed as a non-variable
blazar-like object. However, recently, a low, long-time variability at
$\geq$10\,GeV was reported \citep{1fhl}. No major outburst at
$\gamma$-rays was seen in \pmn, in contrast to most other blazars
\citep{Abdo2010_LC_LBAS}. This lack of short-term variability over
years and across the electromagnetic spectrum is generally unusual,
although \textsl{Fermi}/LAT monitoring shows that blazars can stay in
non-flaring states over long periods \citep{Abdo2010_LC_LBAS}. 
A similar stable behavior of \pmn is seen at radio wavelengths. The
multifrequency ATCA light curve shows no significant variability over
$\sim 4$\,years (see Fig.~\ref{fig:C1730LC}). Note, however, that longterm radio 
monitoring at mm-wavelengths \citep{Hovatta2007} determined 
variability time scales of AGN of about $\sim 4$\,years. 
Archival ATCA data of \pmn going
back about a dozen years, however, show some evidence for low-level,
long timescale flux density variation similar to that seen in some GPS
and CSS sources \citep{Tingay2003}. 
At milliarcsecond resolution, the
TANAMI VLBI observations show no structural or flux density
variability over a period of $\sim$15\,months (see
Sect.~\ref{sec:results}). No fast motion of jet components was
detected, which is uncommon for $\gamma$-ray bright, highly beamed
jets \citep{Lister2011}. However, \citet{Piner2010} found that TeV
blazars can exhibit undetectable parsec-scale motions despite being
very strong in $\gamma$-rays. Note that the TeV blazars are generally
HSPs while \pmn has a synchrotron peak frequency below $10^{14}$\,Hz.
    
\paragraph{Polarization:}

For \pmn, observations by \citet{Murphy2010_AT20G} put a limit on the
overall polarization of $<$1.2\% at 20\,GHz and $<$1\% at 5 and
8.4\,GHz. A higher degree of polarization on smaller angular scales
cannot be excluded and might be detected by polarization measurements
at higher angular resolution, as the VLBI observations presented here
do not include polarimetry.
    
\paragraph{Broadband Spectral Properties:}

The $\gamma$-ray spectral slope of \pmn (see Sect.~\ref{sec:intro})
matches the ones of the luminous, hard-spectrum \textsl{Fermi}/LAT
AGN: \pmn (with
$\Gamma_{100\,\mathrm{MeV}-100\,\mathrm{GeV}}=2.04\pm0.04$) is among
the 30 brightest 2LAC objects \citep{2lac} which have a mean
$\gamma$-ray spectral index of
$\Gamma_{100\,\mathrm{MeV}-100\,\mathrm{GeV}}\approx2.0$. With the
exception of the radio galaxy \object{NGC\,1275} all of these 30
brightest sources are either classified as flat spectrum quasars
(FSRQs) or BL\,Lacs.

Based on its infrared colors, \object{WISE J160350.68$-$490405.6} (the
counterpart associated with \pmn)
would be consistent with the definition of the \textsl{WISE} Gamma-ray Strip for
BL\,Lacs \citep{Massaro2011a, Massaro2012a}. 
However, the ``\textsl{WISE} blazar strip'' is defined based on sources at high
galactic latitudes and all projections in the \textsl{WISE} color space 
were treated separately, while \citet{DAbrusco2013} modeled and selected the 
``\textsl{WISE} locus'' based on the Principal Component space generated by the whole three-dimensional \textsl{WISE}
color distribution corrected for galactic extinction of confirmed $\gamma$-ray blazars.
Using the extinction corrected colors\footnote{Extinction corrected colors are derived from our broadband SED
parametrization (see Sect.~\ref{susec:SED}).} instead and applying the 
method described in \citet{DAbrusco2013}, the source is 
not a candidate blazar within the definition the 3D-\textsl{WISE} locus (R.~D'Abrusco, priv.~comm.).

In a typical BL\,Lac, the non-thermal jet emission would outshine all
other broadband emission components and the infrared component is
expected to be primarily due to synchrotron processes. Contrary to
this, the broadband SED of \pmn shows a strong excess in the infrared
and the broadband parametrization requires an additional blackbody component with
an approximate\footnote{As discussed  Sect.~\ref{susec:confusion}, we cannot exclude that the reported 
\textsl{WISE} fluxes of \pmn are contaminated by the emission of a neighboring
source.} temperature of $\approx 1600$\,K.
Such infrared emission features are not expected to show up in BL\,Lac
SEDs \citep{Abdo2010e}.

The broadband SED parametrization (Fig.~\ref{fig:SED}) shows a striking Compton dominance,
the high energy emission hump is about two orders of magnitude 
above the synchrotron peak, which further challenges the BL\,Lac 
classification scenario\footnote{However, note that the peak flux of 
the  simple, analytical parametrization with two logarithmic parabolas 
is not well constrained 
in the low energy regime.}.
The double-humped broadband emission of BL\,Lacs is generally well described
by synchrotron-self Compton models \citep[e.g.,][]{Ghisellini2010}, and such a dominance at high energies
would then lead to the ``Compton catastrophe''.
As long as no external radiation fields are considered, this particular 
broadband feature of \pmn is difficult to reconcile with the BL\,Lac classification.
Broadband emission models including external photons from the accretion disk,
the broad line region or the cosmic microwave background
\citep[e.g.,][]{Dermer2009,Finke2013,Potter2013} could explain the SED, 
but are more commonly used for flat spectrum radio quasars, and would hence 
be disfavored for this source due to the lack of optical emission lines or 
thermal disk emission.

\paragraph{Milliarcsecond Structure:}

Our TANAMI dual-frequency VLBI observations reveal a resolved and
symmetric brightness distribution on milliarcsecond scales. The
brightest component in the center shows a flat spectrum and the
highest brightness temperature. This result is in contrast to the
typical BL\,Lac structure of an optically thick core component as the
bright end of an one-sided jet, where the spectral index changes from
flat or inverted to steep further downstream
\citep[e.g.,][]{Taylor2005}. The brightness temperature of the three
components of $T_\mathrm{B}\leq 10^{10}$\,K is much lower than
comparable values of blazars, i.e., of beamed emission
\citep{Ojha2010a}.

Due to the limited $(u,v)$-coverage at 22.3\,GHz, no unique
well-defined model of the brightness distribution at this frequency
could be derived. As described above, we tested different extreme
models and image alignments within the ranges allowed by the
visibility data. No other alignment resulted in a physical
representation of the spectral index distribution along the jet (see
Sect.~\ref{susec:spix}).

The comparison of our TANAMI results with flux density measurements of
ATCA reveals that the VLBI observations miss $\sim$20\,\% of the total
flux density at 8.4\,GHz. If \pmn is indeed a BL\,Lac object, this
difference could be explained by halo emission equivalent to the
top-view of FR~I lobes \citep[e.g.,][]{Giroletti2004, Kharb2010} which
might be detected with deeper ATCA observations.

In conclusion, the variability, polarization, and its broadband SED,
as well as the VLBI structure and spatial spectral distribution of
\pmn appears very atypical compared to blazar properties.

\subsection{Alternative Classification Scenarios}\label{susec:alternative}

The peculiarity of \pmn regarding its classification as a BL\,Lac type
object leaves room for alternative interpretations of the
multiwavelength data. In the following we discuss several possible
scenarios in an attempt to get a better description of the observed
broadband properties of \pmn.

\paragraph{\pmn as a Galactic source:}

The source location close to the Galactic plane, the lack of a
redshift measurement, and its non-thermal broadband emission might
suggest a possible Galactic origin. The very low variability at high
energies, however, combined with the lack of periodicity, and the
non-detection of an optical galactic companion excludes the
possibility that \pmn is a black hole binary system. We also
considered the unlikely classification as a pulsar wind nebula (PWN)
even though no matching pulsar is known. Based on its VLBI properties
we can easily exclude this scenario: the radio luminosity of \pmn is
much higher than usually measured for PWNe \citep[e.g.,][]{Frail1997}.
Moreover, assuming that the VLBI structure of $\sim$15\,mas
corresponds to an extension of a few parsec, which is the mean PWN
extension \citep{Gaensler2006}, would place \pmn well outside the
Galaxy at a distance of a few Mpc.
    
\paragraph{Symmetric mas-scale morphology caused by gravitational lensing:}

As \pmn shows a striking symmetry on mas scales, both in structure and
flux density distribution, the possibility of seeing a gravitationally
lensed object should be considered. Assuming that this symmetric VLBI
structure is influenced by lensing causing fainter doubles of the
lensed source within a region of $\sim$15\,mas, the corresponding
Einstein radius would be on the order of several milliarcseconds. It
could also be larger, assuming that the lens is located north/south of
the source, i.e., not the full ring is seen but only a part of it
(yielding the observed elongation). Due to the constant and stable
multiwavelength behavior over months to years, we can exclude a lens
of a few solar masses in the parsec regime. Moreover,
\citet{Wilkinson2001} found that lensing by intergalactic massive
objects in the mas-regime is highly unlikely. The lensed system with
the smallest separation known to date is \object{S3\,0218+35} with
335\,mas \citep{Biggs1999}, while other gravitational lensed systems
are beyond VLBI scales. While we cannot fully exclude a more massive
object ($M_\mathrm{lens} \gtrsim 10^6\,M_\odot$) located at a distance
of a few Mpc causing lensing on mas-scales based on our data, we
consider it to be very unlikely due to the non-detection of a bright
optical source within the line of sight. More exotic lenses with
masses around $10^{2\mbox{--}5}M_\odot$ at shorter distances (such as
intermediate-mass black holes in the Galactic halo or beyond) are
unlikely.

\paragraph{\pmn as an edge-on jet system.}

The 8.4\,GHz VLBI structure of \pmn suggests a double-sided source
like in FRI radio galaxies, but on smaller scales, i.e., a
jet-counterjet system seen side-on. The intrinsic symmetry and
especially the compactness with a small angular separation could
indicate a young, not yet evolved, radio galaxy. In the following we
discuss in particular how the multiwavelength data of \pmn are in
agreement with general properties of Compact Symmetric Objects (CSOs).

The similarity to the double-sided morphologies of present-day radio
galaxies has thus led to the hypotheses that these objects are either
the young versions of their larger FR\,I/FR\,II counterparts
\citep{Fanaroff1974, Readhead1996b, Fanti1995} or ``frustrated'' jets,
i.e., confined due to interaction with a dense, surrounding medium
\citep[e.g.,][]{Bicknell1997, Carvalho1998}. However, the
``evolution'' or ``youth'' scenario is more commonly accepted and
further supported by kinematic measurements of the hot spots revealing
characteristic ages less than $10^{3}$\,years
\citep[e.g.,][]{Owsianik1998,An2012a}. Contrary to flat blazar
spectra, the radio spectra of these compact objects show a peak at
$\sim$1\,GHz or even a steep spectrum, leading to classification as a
Gigahertz-peaked source (GPS) or compact steep spectrum source (CSS),
respectively. CSOs are characterized by a distinct double structure
with an intensity ratio of less than 10:1 and similar steep spectra
and a central, flat-spectrum component \citep{Peck2000,
  Sokolovsky2011b}. They can be distinguished from other radio-loud
AGN based on their low radio variability \citep[typically $<$10\% over
timescales of years; ][]{Fassnacht2001} and their low radio
polarization \citep[$\leq$1\%; ][]{Peck2000}. Due to the high jet
inclination angle little to no motion of individual jet components is
observed \citep[with very few exceptions;][]{Owsianik1998,Taylor2009}.
Since short-time variability is linked to relativistic beaming, we
expect that young radio galaxies show at most only mild broadband
variability.

The possibility of $\gamma$-ray emission from CSO objects has been
considered due to inverse-Compton up-scattering of surrounding photon
fields by the lobe electrons \citep{Stawarz2008,Orienti2011a} and
based on that, the first set of \textsl{Fermi}/LAT sources has been
investigated \citep{McConville2011}, but no unambiguous detection of a
$\gamma$-ray loud CSO has been confirmed, yet. As a class, CSOs are
important sources to probe the origin of high-energy emission in
active galaxies and to study the interaction of jets with the ambient
medium in the nuclear region.

The high-resolution TANAMI images of \pmn at 8.4\,GHz show mas-scale
structure in agreement with a CSO morphology. This structure, the low
$T_\mathrm{B}$ values, and the steep spectral index distribution of
the outer components may indicate a very high jet inclination angle.
In such a case, at most moderate relativistic beaming is expected.
This possible classification is further supported by only mild
long-time multiwavelength variability (Fig.~\ref{fig:C1730LC}), the
lack of apparent superluminal motion and high-energy flares, and the
stability in flux density and size of the individual jet components.
The morphologies of most CSOs show a relatively faint core with
respect to the lobe emission \citep{Taylor1996}. If we see \pmn at a
large inclination angle, the prominent and putative most active
feature would be the central component, which could explain the
observed mild variability. Usually CSOs are dominated by their steep
spectrum lobes but the greater contribution from the core with its
flatter spectrum here could be flattening the overall radio spectrum,
resulting in the observed spectral index of $\alpha = -0.4$. In
particular, the detected brightness temperatures and the overall radio
spectral index of \pmn is consistent with mean values for CSOs of
$T_\mathrm{B, mean} \sim 10^9$\,K and $\alpha_\mathrm{mean} = -0.52$
(between 2.3\,GHz and 8.4\,GHz) reported by \citet{Sokolovsky2011b},
lower than values for bright beamed core-jet sources
\citep{Kovalev2005, Ojha2010a}. Due to very sparse frequency coverage
below 1\,GHz, our ATCA observations are not sufficient to probe a
possible spectral turnover as seen in CSO/GPS sources, though the
archival MGPS data could indicate a GHz-peak. The archival
measurements by \cite{Murphy2010_AT20G} give upper limits to the
overall polarization of $\leq$1.2\% further supporting the CSO
classification.

Our kinematic analysis of three VLBI observation epochs only results
in an upper limit on the apparent jet speed, requiring further
monitoring to check for significant motion in the opposite direction
with respect to the central feature in order to determine the age of
the radio source. However, assuming axis symmetry and a jet
orientation close to the plane of the sky and using the modelfit
parameters (see Table~\ref{table:2}), the flux ratio of the two
extended features gives $R\approx 1.2$, i.e., a corresponding
jet-to-counterjet surface brightness ratio. Hence, jet and counterjet
exhibit a similar Doppler factor. For other sources in the plane of
the sky, e.g., NGC\,1052 \citep{Vermeulen2003}, a pattern speed of
$\beta \approx 0.25$ is measured. Together with the assumption of
optically thin emission with $\alpha = - 0.5$ and using
\begin{equation}\centering
    R = \left(\frac{1+\beta \cos(\theta)}{1-\beta \cos(\theta)}\right)^{3-\alpha},
\end{equation}
this results in an inclination angle of $ \theta \sim 84^\circ$. Since
both, $\beta$ and $\alpha$ are rather conservative assumptions, it is
unlikely that the inclination angle is smaller.

The intriguing double-sided, symmetric 8.4\,GHz emission is confined
to an extension of less than 15\,mas. In standard cosmology
($H_0=73\,\mathrm{km}\,\mathrm{s}^{-1}\,\mathrm{Mpc}^{-1}$,
$\Omega_\mathrm{m}=0.27$, $\Omega_\lambda=0.73$), the
redshift-dependent ratio between angular size and linear extension
peaks at $z\sim1.6$. As a consequence the linear diameter of \pmn
cannot be larger than $d_\mathrm{limit} \lesssim 125$\,pc, placing the
VLBI resolved structure well below the theoretical limit of CSO
extensions. There is evidence for a western extension of the mas-scale
jet seen in the 8.4\,GHz VLBI data and we find a discrepancy between
TANAMI and ATCA flux density measurements at 8.4\,GHz of $\sim$20\%.
This could indicate radio emission at larger scales. Although
unexpected for young radio sources, such extended emission in
association with a young double source has been reported in the
literature \citep{Baum1990,Tingay2003} and proposed to be due to
recurrent activity in these objects with the extended emission
attributed to past epochs of activity. \citet[][their
Fig.~9]{Edwards2004} suggest that such `missing' flux between ATCA and
VLBI baselines is more common for a CSS than for a GPS source. ATCA
observations at 5.5\,GHz and 9\,GHz show no extended emission,
confirming the compact overall structure of this source down to~$\sim
1'$.

\paragraph{Starburst-like broadband emission:}
In general the spectral energy distribution resembles the broadband
emission of starburst galaxies \citep{Lenain2010}, where the enhanced
IR emission is due to enhanced star formation activity. One expects
temperatures on the order of a few tens of Kelvin for dust emission
\citep{Elbaz2011} while higher temperatures are explained by enhanced
star emission. Nuclear temperatures around \mbox{(1--2)$\times
  10^3$\,K} are found in Type~II Seyferts \citep{Lira2013}.

Four starburst galaxies are also detected in $\gamma$-rays by
\textsl{Fermi}/LAT \citep{Abdo2010_starburst, 2lac}, \object{M\,82}
and \object{NGC 253} even up into the TeV range
\citep{Veritas2009_M82, HESS2009_NGC253}. For this source type, the
origin of the $\gamma$-ray emission is explained by interactions of
cosmic rays with interstellar gas rather than due to AGN activity. At
these high energies, starburst galaxies show no significant
variability with spectral indices in the range of $\Gamma_\gamma =
2.1\mbox{--}2.4$ \citep{Ackermann2012_starburst}. In contrast, \pmn
shows a mild variability on longer timescales in the $\gamma$-rays
\citep{1fhl}.

Moreover, the $\gamma$-ray fluxes of these starburst galaxies are in
the range of $(5\mbox{--}11)\times
10^{-10}\,\mathrm{ph}\,\mathrm{cm}^{-2}\,\mathrm{s}^{-1}$
\citep{2lac}. These fluxes are about ten times fainter than the
reported $\gamma$-ray flux of \pmn, which is the most important caveat
for this interpretation. Assuming a similar $\gamma$-ray luminosity
for all sources of this class and assuming the $\gamma$-ray emission
of \pmn is only due to starburst activity, this would give a distance
approximately three times closer than these $\gamma$-bright starburst
galaxies ($\sim$1--4\,Mpc, comparable to the Andromeda galaxy or
\object{Centaurus~A}). Despite high extinction and high source density
along this line of sight, it is very unlikely that such a close object
would have been missed by multiwavelength observations until now. This
constraint is a strong argument against the starburst interpretation
of the SED, but only if no $\gamma$-rays are emitted by an additional
AGN or jet. These constraints require further statistics on the
$\gamma$-ray properties of starbursts.

\cite{Lenain2010} report on the $\gamma$-ray emission of radio-loud
starburst galaxies and conclude that the parsec-scale jet of
\object{NGC\,1068} shows a significant contribution to the high-energy
emission. These findings could explain the long-term $\gamma$-ray
variability and the higher flux compared to other $\gamma$-ray bright
starbursts by a significant influence of the AGN. This is further
supported by the high core dominance, suggesting that the center is
the putative most active region if we really see the source at a large
inclination angle, as discussed in the previous paragraph.
 
Though better multiwavelength coverage is required to precisely
interpret the SED, it is striking that the IR emission, the
$\gamma$-ray spectral index and its low (long-term) variability are in
good agreement with starburst broadband properties. If further optical
and IR observations confirm starburst activity, then \pmn would be the
$\gamma$-ray brightest object in this particular source class.

Observations of radio-loud starbursts show evidence for a
CSO-starburst connection. \cite{Tadhunter2011} find that radio-loud
starburst galaxies tend to show morphologies unlike typical
FR\,I/II-class objects. Compact radio sources ($d < 15$\,kpc) are
found to have a significantly higher star formation rate than more
evolved sources, however, a possible selection effect must be
considered \citep{Dicken2012}. Hence, no simple interpretation of
merger-induced star formation triggering jet activity can generally be
applied.

According to the available multiwavelength data and keeping in mind
the most limiting factor, the lack of well constrained optical
spectra, this classification scenario gives an intriguing picture of
the possible nature of \pmn. The $\gamma$-ray emission could be
explained by a contribution of the starburst activity but definitely
requires an additional (dominating) spectral component to model the
contribution of the AGN emission to compensate for the disagreement in
brightness and long-term variability not seen in $\gamma$-ray bright
starburst galaxies. This alternative interpretation of the broadband
emission would be more plausible, but contributions of the young radio
jet as proposed by \cite{Stawarz2008} should still be considered.
%
\section{Summary and Conclusions}\label{sec:conclusions}

We have presented the first VLBI observations and additional
multiwavelength data for the $\gamma$-ray bright source \pmn. It
exhibits multiwavelength properties atypical for well-established
$\gamma$-ray emitting object classes. We have discussed several
classification scenarios and found that the source is most likely
either a very peculiar BL\,Lac object or a CSO.
The $\gamma$-ray brightness is well explained if \pmn is indeed a
blazar, though the broadband emission and the mas-scale structure are
very unusual. These spectral and structural features, atypical for a
blazar, are accounted for by the alternative classification as a
misaligned young radio source, possibly in a host galaxy with
starburst activity.

If it is confirmed that this source is indeed not a BL\,Lac, it will
add to the class of misaligned $\gamma$-ray bright sources. Only a few
such sources are known and \pmn would be the first CSO in this class
\citep{Abdo2010_misaligned}. The confirmed detection of a $\gamma$-ray
bright CSO would challenge current jet emission models attempting to
explain the high-energy spectral component with high beaming factors
and would help to determine the region from where $\gamma$-rays are
emitted. Hence, \pmn would be an important object to study the emission of
high-energy photons in these misaligned sources and to address the
open question of whether their $\gamma$-ray brightness is due to a
beamed jet, lobe emission, or a combination of both.

TANAMI observations are ongoing, and with additional VLBI epochs, the
kinematic analysis will be better constrained. Low resolution radio
observations will be made at lower frequencies to better constrain the
radio spectrum.

We will be attempting deeper photometric observations in the NIR
(where the effects of Galactic extinction would be less severe) in
order to establish the point source counterpart and obtain a spectrum
further constraining the redshift. With additional X-ray and optical
observations, we will be able to better model and constrain the
broadband SED.

\begin{acknowledgements}
  The Australian Long Baseline Array and the Australia Telescope
  Compact Array are part of the Australia Telescope National Facility
  which is funded by the Commonwealth of Australia for operation as a
  National Facility managed by CSIRO. This research was funded in part
  by NASA through {\em Fermi} Guest Investigator grants NNH09ZDA001N
  and NNH10ZDA001N. It was supported by an appointment to the NASA
  Postdoctoral Program at the Goddard Space Flight Center,
  administered by Oak Ridge Associated Universities through a contract
  with NASA. It is based on observations obtained at the Gemini
  Observatory (Program ID: GS-2013A-Q-80), which is operated by the
  Association of Universities for Research in Astronomy, Inc., under a
  cooperative agreement with the NSF on behalf of the Gemini
  partnership: the National Science Foundation (United States), the
  National Research Council (Canada), CONICYT (Chile), the Australian
  Research Council (Australia), Minist\'erio da Ci\^encia, Tecnologia
  e Inova\c{c}\~{a}o (Brazil) and Ministerio de Ciencia, Tecnolog\'ia
  e Innovaci\'on Productiva (Argentina). This research has made use of
  data from the NASA/IPAC Extragalactic Database (NED), operated by
  the Jet Propulsion Laboratory, California Institute of Technology,
  under contract with the National Aeronautics and Space
  Administration; and the SIMBAD database (operated at CDS,
  Strasbourg, France).
  
  C.M.~acknowledges the support of the `Studienstiftung des Deutschen
  Volkes'. E.R.~was partially supported by the Spanish MINECO projects
  AYA2009-13036-C02-02 and AYA2012-38491-C02-01 and by the Generalitat
  Valenciana project PROMETEO/2009/104, as well as by the COST MP0905
  action `Black Holes in a Violent Universe'. R.S.\ acknowledges the
  support of Deutsche Forschungsgemeinschaft grant DFG WI1860/10-1. We
  thank I.~Donnarumma for the useful discussion about starburst SEDs, 
  R.~D'Abrusco for performing checks on the corrected \textsl{WISE} data,
  and J.E.~Davis for the development of the \texttt{slxfig} module
  used to prepare all figures in this work.

  We thank the referee for the helpful comments 
  and suggestions.
  
\end{acknowledgements}
\bibliographystyle{aa} 
\bibliography{mnemonic,aaabbrv,pmnj1603-4904}
%
 
\end{document}